\documentclass[11pt,reqno]{amsart}

\usepackage{amscd,amssymb,amsmath,amsthm}
\usepackage{graphicx}
\usepackage{color}
\usepackage{cite}
\usepackage{epstopdf}
\usepackage{amssymb}
\usepackage[russian]{babel}
\numberwithin{equation}{section} 

\def\s{\sigma}

\def\s{\sigma}

\def\s{\sigma}

\begin{document}
\begin{center}
\textbf{\Large {╥Ёрэёы Ўшюээю-шэтрЁшрэЄэ√х ш яхЁшюфшўхёъшх ьхЁ√ ├шссёр фы  ьюфхыш ╧юЄЄёр эр фхЁхтх ╩¤ыш}}\\
\end{center}\

\begin{center}
╨.╠.╒ръшьют\footnote{╚эёЄшЄєЄ ьрЄхьрЄшъш, єы. ─єЁьюэ щєыш, 29, ╥р°ъхэЄ, 100125, ╙чсхъшёЄрэ.\\
E-mail: rustam-7102@rambler.ru}, ╘.╒.╒рщфрЁют\footnote{═рЎшюэры№э√щ єэштхЁёшЄхЄ ╙чсхъшёЄрэр, ╥р°ъхэЄ, ╙чсхъшёЄрэ.\\
E-mail: haydarov\_imc@mail.ru},
\end{center}\

─ы  ЇхЁЁюьруэшЄэющ ьюфхыш ╧юЄЄёр ё ЄЁхь  ёюёЄю эш ьш шчєўхэ√ ЄЁрэёы Ўшюээю-шэтрЁшрэЄэ√х ьхЁ√ ├шссёр эр фхЁхтх ╩¤ыш яюЁ фър $k=3$ ш фрэ√  тэ√х ЇюЁьєы√ фы  ЄЁрэёы Ўшюээю-шэтрЁшрэЄэ√ї ьхЁ ├шссёр. ╚чєўр■Єё  яхЁшюфшўхёъшх ьхЁ√ ├шссёр фы  рэЄшЇхЁЁюьруэшЄэющ ьюфхыш ╧юЄЄёр ё $q-$ёюёЄю эш ьш эр фхЁхтх ╩¤ыш яюЁ фър $k$. ╧Ёш ¤Єюь єыєў°хэ√ эхъюЄюЁ√х Ёхчєы№ЄрЄ√ шч ЁрсюЄ√ \cite{KhR}: эр эхъюЄюЁ√ї шэтрЁшрэЄрї фрэю Єюўэюх ъюышўхёЄтю яхЁшюфшўхёъшї ьхЁ ├шссёр эр фхЁхтх ╩¤ыш яюЁ фър $k\geq3$ ё яхЁшюфюь фтр. \\

\textbf{╩ы■ўхт√х ёыютр}: фхЁхтю ╩¤ыш, ъюэЇшуєЁрЎш , ьюфхы№ ╧юЄЄёр, ьхЁр ├шссёр, ЄЁрэёы Ўшюээю-шэтрЁшрэЄэ√х ьхЁ√,
 яхЁшюфшўхёъшх ьхЁ√.

\section{\textbf{┬тхфхэшх}}\

╚чтхёЄэю, ўЄю ърцфющ яЁхфхы№эющ ьхЁх ├шссёр ёюяюёЄрты хЄё  юфэр Їрчр Їшчшўхёъющ ёшёЄхь√. ╧ю¤Єюьє т ЄхюЁшш ьхЁ ├шссёр юфэющ шч трцэ√ї чрфрў  ты хЄё  ёє∙хёЄтютрэшх Їрчютюую яхЁхїюфр, Є.х. ъюуфр Їшчшўхёър  ёшёЄхьр ьхэ хЄ ётюх ёюёЄю эшх яЁш шчьхэхэшш ЄхьяхЁрЄєЁ√. ▌Єю яЁюшёїюфшЄ, ъюуфр ьхЁр ├шссёр эх хфшэёЄтхээр. ╧Ёш ¤Єюь ЄхьяхЁрЄєЁр, яЁш ъюЄюЁющ ьхэ хЄё  ёюёЄю эшх Їшчшўхёъющ ёшёЄхь√, юс√ўэю эрч√трхЄё  ъЁшЄшўхёъющ (\cite {6}-\cite {Si}).

┬ \cite{Ga8} шчєўхэр ЇхЁЁюьруэшЄэр  ьюфхы№ ╧юЄЄёр ё ЄЁхь 
ёюёЄю эш ьш эр фхЁхтх ╩¤ыш тЄюЁюую яюЁ фър ш яюърчрэю
ёє∙хёЄтютрэшх ъЁшЄшўхёъющ ЄхьяхЁрЄєЁ√ $T_c$ Єръющ, ўЄю яЁш $T<T_c$
ёє∙хёЄтє■Є ЄЁш ЄЁрэёы Ўшюээю-шэтрЁшрэЄэ√ї ш эхёўхЄэюх ўшёыю эх
ЄЁрэёы Ўшюээю-шэтрЁшрэЄэ√ї ьхЁ ├шссёр. ┬ ЁрсюЄх \cite {GN}
юсюс∙хэ√ Ёхчєы№ЄрЄ√ ЁрсюЄ√ \cite{Ga8} фы  ьюфхыш ╧юЄЄёр ё ъюэхўэ√ь
ўшёыюь ёюёЄю эшщ эр фхЁхтх ╩¤ыш яЁюшчтюы№эюую (ъюэхўэюую) яюЁ фър.

┬ ЁрсюЄх \cite{R} яюърчрэю, ўЄю эр фхЁхтх ╩¤ыш
ЄЁрэёы Ўшюээю-шэтрЁшрэЄэр  ьхЁр ├шссёр рэЄшЇхЁЁюьруэшЄэющ ьюфхыш
╧юЄЄёр ё тэх°эшь яюыхь хфшэёЄтхээр.
╨рсюЄр \cite{Ga13} яюёт ∙хэр ьюфхыш ╧юЄЄёр ёю ёўхЄэ√ь ўшёыюь ёюёЄю эшщ ш
c эхэєыхт√ь тэх°эшь яюыхь ш фюърчрэю, ўЄю ¤Єр ьюфхы№ шьххЄ
хфшэёЄтхээє■ ЄЁрэёы Ўшюээю-шэтрЁшрэЄэє■ ьхЁє ├шссёр.

┬ ЁрсюЄх \cite{RK} шчєўхэ√ яхЁшюфшўхёъшх ьхЁ√ ├шссёр ш яЁш эхъюЄюЁ√ї єёыютш ї фюърчрэю, ўЄю тёх
яхЁшюфшўхёъшх ьхЁ√ ├шссёр  ты ■Єё  ЄЁрэёы Ўшюээю-шэтрЁшрэЄэ√ьш;
эрщфхэ√ єёыютш , яЁш ъюЄюЁ√ї ьюфхы№ ╧юЄЄёр ё эхэєыхт√ь тэх°эшь
яюыхь шьххЄ яхЁшюфшўхёъшх ьхЁ√ ├шссёр. ╨рсюЄр \cite{KhR1}  ты хЄё  яЁюфюыцхэшхь ЁрсюЄ√ \cite{RK}.
─юърчрэр ёє∙хёЄтютрэшх эх ьхэхх ЄЁхї яхЁшюфшўхёъшї ьхЁ ├шссёр ё яхЁшюфюь фтр эр фхЁхтх ╩¤ыш яюЁ фър
ЄЁш ш ўхЄ√Ёх фы  ьюфхыш ╧юЄЄёр ё ЄЁхь  ёюёЄю эш ьш ш ё эєыхт√ь тэх°эшь
яюыхь. └ т ЁрсюЄх \cite{KhR} шчєўхэр ьюфхы№ ╧юЄЄёр ё $q-$ёюёЄю эш ьш
эр фхЁхтх ╩¤ыш яюЁ фър $k\geq3$ ш эр эхъюЄюЁ√ї шэтрЁшрэЄрї яюърчрэю ёє∙хёЄтютрэшх яхЁшюфшўхёъшї
(эх ЄЁрэёы Ўшюээю-шэтрЁшрэЄэ√ї) ьхЁ ├шссёр яЁш эхъюЄюЁ√ї єёыютш ї
эр ярЁрьхЄЁ√ ¤Єющ ьюфхыш. ╩Ёюьх Єюую, єърчрэр эшцэ   уЁрэшЎр
ъюышўхёЄтр ёє∙хёЄтє■∙шї яхЁшюфшўхёъшї ьхЁ ├шссёр. ┬ ЁрсюЄх \cite{KRK} фрэю яюыэюх юяшёрэшх ЄЁрэёы Ўшюээю-шэтрЁшрэЄэ√ї ьхЁ ├шссёр
фы  ЇхЁЁюьруэшЄэющ ьюфхыш ╧юЄЄёр ё $q-$ёюёЄю эш ьш ш яюърчрэю, ўЄю
шї ъюышўхёЄтю Ёртэю $2^q-1$, р т ЁрсюЄх \cite{KR} шчєўхэр чрфрўр ъЁрщэюёЄш ¤Єшї ьхЁ.

┬ ¤Єющ ЁрсюЄх ь√ фрфшь  тэ√х ЇюЁьєы√ фы  ЄЁрэёы Ўшюээю-шэтрЁшрэЄэ√ї ьхЁ ├шссёр ЇхЁЁюьруэшЄэющ ьюфхыш ╧юЄЄёр ё ЄЁхь  ёюёЄю эш ьш эр фхЁхтх ╩¤ыш яюЁ фър $k=3$. ╩Ёюьх Єюую, фюърцхь, ўЄю эр эхъюЄюЁюь шэтрЁшрэЄх яЁш эхъюЄюЁ√ї єёыютш ї эр ярЁрьхЄЁ√
рэЄшЇхЁЁюьруэшЄэющ ьюфхыш ╧юЄЄёр ё $q-$ёюёЄю эш ьш ё эєыхт√ь тэх°эшь яюыхь эр
фхЁхтх ╩¤ыш яюЁ фър $k\geq 3$ ёє∙хёЄтє■Є Ёютэю фтх
яхЁшюфшўхёъшх (эх ЄЁрэёы Ўшюээю-шэтрЁшрэЄэ√х) ьхЁ√ ├шссёр ё яхЁшюфюь фтр ш єърцхь Єюўэюх ъюышўхёЄтю ёє∙хёЄтє■∙шї яхЁшюфшўхёъшї ьхЁ ├шссёр эр юс·хфшэхэшш эхъюЄюЁ√ї шэтрЁшрэЄэ√ї ьэюцхёЄт. \

\section{\textbf{╬яЁхфхыхэш  ш шчтхёЄэ√х ЇръЄ√}}\

─хЁхтю ╩¤ыш $\Im^k$ яюЁ фър $ k\geq 1 $ - схёъюэхўэюх фхЁхтю, Є.х.
уЁрЇ схч Ўшъыют, шч ърцфющ тхЁ°шэ√ ъюЄюЁюую т√їюфшЄ Ёютэю $k+1$
ЁхсЁю. ╧єёЄ№ $\Im^k=(V,L,i)$, уфх $V-$хёЄ№ ьэюцхёЄтю тхЁ°шэ
$\Im^k$, $L-$ьэюцхёЄтю хую ЁхсхЁ, ш $i-$ЇєэъЎш 
шэЎшфхэЄэюёЄш, ёюяюёЄрты ■∙р  ърцфюьє ЁхсЁє $l\in L$ хую ъюэЎхт√х
Єюўъш $x, y \in V$. ┼ёыш $i (l) = \{ x, y \} $, Єю $x$ ш $y$
эрч√тр■Єё   {\it сышцрщ°шьш ёюёхф ьш тхЁ°шэ√} ш юсючэрўрхЄё  $l =
<x, y> $. ╨рёёЄю эшх $d(x,y), x, y \in V$ эр фхЁхтх ╩¤ыш
юяЁхфхы хЄё  ЇюЁьєыющ
$$
d (x, y) = \min \ \{d | \exists x=x_0, x_1,\dots, x _ {d-1},
x_d=y\in V \ \ \mbox {Єръющ, ўЄю} \ \ <x_0, x_1>,\dots, <x _
{d-1}, x_d>\}.$$

─ы  ЇшъёшЁютрээюую $x^0\in V$ юсючэрўшь $ W_n = \ \{x\in V\ \ | \ \
d (x, x^0) =n \}, $
$$ V_n = \ \{x\in V\ \ | \ \ d (x, x^0) \leq n \},\ \
L_n = \ \{l = <x, y> \in L \ \ | \ \ x, y \in V_n \}. $$

╚чтхёЄэю, ўЄю ёє∙хёЄтєхЄ тчршьэююфэючэрўэюх ёююЄтхЄёЄтшх ьхцфє
ьэюцхёЄтюь $V$ тхЁ°шэ фхЁхтр ╩¤ыш яюЁ фър $k\geq 1 $ ш уЁєяяющ $G
_{k},$  ты ■∙хщё  ётюсюфэ√ь яЁюшчтхфхэшхь $k+1$ Ўшъышўхёъшї уЁєяя
тЄюЁюую яюЁ фър ё юсЁрчє■∙шьш $a_1, a_2,\dots, a_{k+1} $,
ёююЄтхЄёЄтхээю.\

╠√ ЁрёёьюЄЁшь ьюфхы№, уфх ёяшэют√х яхЁхьхээ√х яЁшэшьр■Є чэрўхэш 
шч ьэюцхёЄтр $\Phi = \ \{1, 2,\dots, q \},$ $ q\geq 2 $ ш
Ёрёяюыюцхэ√ эр тхЁ°шэрї фхЁхтр. ╥юуфр \emph{ ъюэЇшуєЁрЎш } $\s$ эр
$V$ юяЁхфхы хЄё  ъръ ЇєэъЎш  $x\in V\to\s (x) \in\Phi$; ьэюцхёЄтю
тёхї ъюэЇшуєЁрЎшщ ёютярфрхЄ ё $\Omega =\Phi ^ {V} $.

├рьшы№Єюэшрэ ьюфхыш ╧юЄЄёр юяЁхфхы хЄё  ъръ
\begin{equation}\label{f.1}
H(\sigma)=-J\sum_{\langle x,y\rangle\in L}
\delta_{\sigma(x)\sigma(y)},
\end{equation}
уфх $J\in R$, $\langle x,y\rangle-$ сышцрщ°шх ёюёхфш ш
$\delta_{ij}-$ ёшьтюы ╩ЁюэхъхЁр:
$$\delta_{ij}=\left\{\begin{array}{ll}
0, \ \ \mbox{хёыш} \ \ i\ne j\\[2mm]
1, \ \ \mbox{хёыш} \ \ i= j.
\end{array}\right.
$$
╬яЁхфхышь ъюэхўэюьхЁэюх ЁрёяЁхфхыхэшх тхЁю ЄэюёЄэющ ьхЁ√ $\mu$ т
юс№хьх $V_n$ ъръ
\begin{equation}\label{f.2}
\mu_n(\sigma_n)=Z_n^{-1}\exp\left\{-\beta
H_n(\sigma_n)+\sum_{x\in W_n}h_{\sigma(x),x}\right\},
\end{equation}
уфх $\beta=1/T$, $T>0$--ЄхьяхЁрЄєЁр,  $Z_n^{-1}$ эюЁьшЁє■∙шщ
ьэюцшЄхы№ ш $\{h_x=(h_{1,x},\dots, h_{q,x})\in R^q, x\in V\}$
ёютюъєяэюёЄ№ тхъЄюЁют ш
$$H_n(\sigma_n)=-J\sum_{\langle x,y\rangle\in L_n}
\delta_{\sigma(x)\sigma(y)}.$$

├ютюЁ Є, ўЄю тхЁю ЄэюёЄэюх ЁрёяЁхфхыхэшх (\ref{f.2}) ёюуырёютрээюх, хёыш
фы  тёхї $n\geq 1$ ш $\sigma_{n-1}\in \Phi^{V_{n-1}}$:
$$\sum_{\omega_n\in \Phi^{W_n}}\mu_n(\sigma_{n-1}\vee
\omega_n)=\mu_{n-1}(\sigma_{n-1}).$$

╟фхё№ $\sigma_{n-1}\vee \omega_n$  хёЄ№ юс·хфшэхэшх ъюэЇшуєЁрЎшщ.
┬ ¤Єюь ёыєўрх, ёє∙хёЄтєхЄ хфшэёЄтхээр  ьхЁр $\mu$ эр $\Phi^V$
Єрър , ўЄю фы  тёхї $n$ ш $\sigma_n\in \Phi^{V_n}$
$$\mu(\{\sigma|_{V_n}=\sigma_n\})=\mu_n(\sigma_n).$$
╥рър  ьхЁр эрч√трхЄё  Ёрё∙хяыхээющ ушссёютёъющ ьхЁющ,
ёююЄтхЄётє■∙хщ урьшы№Єюэшрэє (\ref{f.1}) ш тхъЄюЁчэрўэющ ЇєэъЎшш $h_x,
x\in V$.\

╤ыхфє■∙хх єЄтхЁцфхэшх юяшё√трхЄ єёыютшх эр $h_x$, юсхёяхўштр■∙хх
ёюуырёютрээюёЄ№ $\mu_n(\sigma_n)$.

\textbf{╥хюЁхьр 1}.\cite{R} \textit{┬хЁю ЄэюёЄэюх ЁрёяЁхфхыхэшх
$\mu_n(\sigma_n)$, $n=1,2,\ldots$ т (\ref{f.2})  ты хЄё  ёюуырёютрээ√ь
Єюуфр ш Єюы№ъю Єюуфр}, \textit{ъюуфр фы  ы■сюую} $x\in V$
\textit{шьххЄ ьхёЄю ёыхфє■∙хх
\begin{equation}\label{f.3}
h_x=\sum_{y\in
S(x)}F(h_y,\theta),
\end{equation}
уфх $F: h=(h_1,
\dots,h_{q-1})\in R^{q-1}\to
F(h,\theta)=(F_1,\dots,F_{q-1})\in R^{q-1}$ юяЁхфхы хЄё 
ъръ:
$$F_i=\ln\left({(\theta-1)e^{h_i}+\sum_{j=1}^{q-1}e^{h_j}+1\over
\theta+ \sum_{j=1}^{q-1}e^{h_j}}\right),$$
$\theta=\exp(J\beta)$, $S(x)-$ ьэюцхёЄтю яЁ ь√ї яюЄюьъют Єюўъш
$x$} \textit{ш}
$h_x=\left(h_{1,x},\dots,h_{q-1,x}\right)$ \textit{ё єёыютшхь}
\begin{equation}\label{hh}
h_{i,x}={\tilde h}_{i,x}-{\tilde h}_{q,x}, \ \ i=1,\dots,q-1.
\end{equation} \

╧єёЄ№ $\widehat{G}_k-$ яюфуЁєяяр уЁєяя√ $G_k$.

\textbf{╬яЁхфхыхэшх 1}. ╤ютюъєяэюёЄ№ тхъЄюЁют $h=\{h_x,\, x\in
G_k\}$ эрч√трхЄё  $ \widehat{G}_k$-яхЁшюфшўхёъющ, хёыш
$h_{yx}=h_x$ фы  $\forall x\in G_k, y\in\widehat{G}_k.$

$G_k-$ яхЁшюфшўхёъшх ёютюъєяэюёЄш эрч√тр■Єё 
ЄЁрэёы Ўшюээю-шэтрЁшрэЄэ√ьш.

\textbf{╬яЁхфхыхэшх 2}. ╠хЁр $\mu$ эрч√трхЄё 
$\widehat{G}_k$-яхЁшюфшўхёъющ, хёыш юэр ёююЄтхЄёЄтєхЄ
$\widehat{G}_k$-яхЁшюфшўхёъющ ёютюъєяэюёЄш тхъЄюЁют $h$.

╤ыхфє■°р  ЄхюЁхьр їрЁръЄхЁшчєхЄ яхЁшюфшўхёъшх ьхЁ√ ├шссёр.

\textbf{╥хюЁхьр 2.}\cite{RK} \textit{╧єёЄ№ $K-$ эюЁьры№э√щ
фхышЄхы№ ъюэхўэюую шэфхъёр т $G_k.$ ╥юуфр фы  ьюфхыш ╧юЄЄёр тёх
$K-$ яхЁшюфшўхёъшх ьхЁ√ ├шссёр  ты ■Єё  ышсю $G_k^{(2)}-$
яхЁшюфшўхёъшьш, ышсю ЄЁрэёы Ўшюээю-шэтрЁшрэЄэ√ьш, уфх
$G^{(2)}_k=\{x: |x|-\emph{ўхЄэр }\}.$}\

\section{\textbf{╥Ёрэёы Ўшюээю-шэтрЁшрэЄэ√х ьхЁ√ ├шссёр}}\

─ы  ы■сюую $x\in V$ ЄЁрэёы Ўшюээю-шэтрЁшрэЄэр  ьхЁр ├шссёр ёююЄтхЄёЄтєхЄ Ёх°хэш■ $h_x$ шч ЇюЁьєы√ (\ref{f.3}) ё
$h_x=h=(h_1,\dots,h_{q-1})\in R^{q-1}$. ╥юуфр шч єЁртэхэш  (\ref{f.3}) ь√ яюыєўшь $h=kF(h,\theta)$ ш юсючэрўр 
$z_i=\exp(h_i), i=1,\dots,q-1$, яюёыхфэхх єЁртэхэшх ьюцхь яхЁхяшёрЄ№ ъръ
\begin{equation}\label{pt1}
z_i=\left({(\theta-1)z_i+\sum_{j=1}^{q-1}z_j+1\over \theta+
\sum_{j=1}^{q-1}z_j}\right)^k,\ \ i=1,\dots,q-1.
\end{equation}

╚ч ЁрсюЄ√ \cite{KRK} шчтхёЄэю, ўЄю ы■ср  ЄЁрэёы Ўшюээю-шэтрЁшрэЄэр  ьхЁр ╧юЄЄё ьюфхыш ёююЄтхЄёЄтєхЄ Ёх°хэш■ ёыхфє■∙хую єЁртэхэш 
\begin{equation}\label{rm}
z=f_m(z)\equiv \left({(\theta+m-1)z+q-m\over
mz+q-m-1+\theta}\right)^k,
\end{equation}
фы  эхъюЄюЁ√ї $m=1,\dots,q-1$.

╨рёёьюЄЁшь (\ref{rm}) яЁш $k=3, \ \theta>1$ ш юсючэрўшь $\sqrt[3]{z}=x$. ╥юуфр
$$mx^4-(\theta+m-1)x^3+(q-m-1+\theta)x-q+m=0.$$
╟рьхЄшь, ўЄю $x_0=1$  ты хЄё  Ёх°хэшхь ¤Єюую єЁртэхэш . ╨рчфхышт ¤Єю єЁртэхэшх эр $(x-1)$, яюыєўшь
\begin{equation}\label{3.1}
\varphi(x)=mx^3-(\theta-1)x^2-(\theta-1)x+q-m=0.
\end{equation}
╟рьхЄшь, ўЄю яюёыхфэхх єЁртэхэшх яю ЄхюЁхьх ─хърЁЄр ю ъюышўхёЄтх яюыюцшЄхы№э√ї ъюЁэхщ ьэюуюўыхэр шьххЄ эх сюыхх фтєї яюыюцшЄхы№э√ї Ёх°хэш . ╩Ёюьх Єюую, $\varphi(0)=q-m>0, \
\varphi(+\infty)=+\infty$ ш єЁртэхэшх $\varphi'(x)=0$ шьххЄ
хфшэёЄтхээюх яюыюцшЄхы№эюх Ёх°хэшх:
$$x^*(\theta,m)={\theta-1+\sqrt{(\theta-1)^2+3m(\theta-1)}\over 3m},$$
Є.х. ЇєэъЎш  $\varphi(x)$ єс√трхЄ яЁш $x<x^*$ ш тючЁрёЄрхЄ яЁш
$x>x^*$. ╤ыхфютрЄхы№эю, ёє∙хёЄтєхЄ $\theta_{cr}(m,q)$ Єръюх, ўЄю
єЁртэхэшх (\ref{3.1}) эх шьххЄ Ёх°хэш  яЁш $\theta<\theta_{cr}(m,q)$,
шьххЄ юфэю Ёх°хэшх яЁш $\theta=\theta_{cr}(m,q)$ ш шьххЄ фтр Ёх°хэш 
яЁш $\theta>\theta_{cr}(m,q)$, уфх $\theta_{cr}(m,q)$ хёЄ№ Ёх°хэшх єЁртэхэш 
$\varphi(x^*)=0$.

┬ююс∙х уютюЁ , т ЁрсюЄх \cite{KRK} фрэ√ ЇюЁьєы√ (3.17), (3.18) фы  т√ўшёыхэш  $\theta_{cr}(m,q)=\theta_m$ ш шёяюы№чє  ¤Єш ЇюЁьєы√ яЁш $k=3$, шьххь

\begin{equation}\label{3.3}
\theta_{cr}(m,q)=\psi(x^{**}),
\end{equation}
уфх
$$\psi(x)={mx^3+q-m\over x^2+x}+1$$
ш $x^{**}$ хёЄ№ Ёх°хэшх єЁртэхэш  $mx^4+2mx^3-2(q-m)x-(q-m)=0.$ ╨х°шь яюёыхфэхх єЁртэхэшх ёЄрэфрЁЄэ√ь ьхЄюфюь ╘хЁЁрЁш шч ышэхщэющ рыухсЁ√. ╥юуфр яюёых уЁюьючфъшї т√ўшёыхэшщ сєфхь шьхЄ№
$$x^{**}={\sqrt[4]{8\alpha_0^3}+\sqrt{(3-2\alpha_0)\sqrt{2\alpha_0}-6+{4q\over m}}\over 2\sqrt[4]{2\alpha_0}}-{1\over2},$$
уфх
$$\alpha_0={\sqrt[3]{m(8m^2-12mq+4q^2)}+m\over 2m}.$$

\textbf{╤ыєўрщ $q=3$.} ┬ ¤Єюь ёыєўрх єЁртэхэшх $\varphi'(x)=0$ шьххЄ
хфшэёЄтхээюх яюыюцшЄхы№эюх Ёх°хэшх:
$$x^*(\theta,1)={\theta-1+\sqrt{\theta^2+\theta-2}\over 3}$$
ш $\theta_{cr}$ хёЄ№ Ёх°хэшх єЁртэхэш 
$\varphi(x^*)=0$: $\theta_{cr}=\sqrt{9+6\sqrt{3}}-2\approx2.403669476$.

\textbf{╟рьхўрэшх 1.} ▌Єю чэрўхэшх фы  $\theta_{cr}$ ьюцэю яюыєўшЄ№ шч ЇюЁьєы√ (\ref{3.3}) яЁш $q=3, m=1$.

═рщфхь Ёх°хэш  єЁртэхэш  (\ref{3.1}) ё яюью∙№■ ЇюЁьєы√ ╩рЁфрэю. ─ы  ¤Єюую
юсючэрўшь $x=y+(\theta-1)/3$ ш яхЁхяш°хь єЁртэхэшх (\ref{3.1})
\begin{equation}\label{3.2}
y^3+py+r=0,
\end{equation}
уфх
$$p=-{1\over3}(\theta^2+\theta-2), \ r={1\over27}\left(-2(\theta-1)^3-9(\theta-1)^2+54\right).$$
╧юёыхфэхх єЁртэхэшх шьххЄ юфэю юЄЁшЎрЄхы№эюх Ёх°хэшх
$$y=\sqrt[3]{-{r\over2}+\sqrt{{r^2\over4}+{p^3\over27}}}+\sqrt[3]{-{r\over2}-\sqrt{{r^2\over4}+{r^3\over27}}}$$
яЁш $\theta<\theta_{cr}$, юфэю яюыюцшЄхы№эюх Ёх°хэшх
$$y^*={1\over2}\left(1+\sqrt{3}+{1\over3}\sqrt{9+6\sqrt{3}}(1-\sqrt{3})\right)\approx 0.828740438$$
яЁш $\theta=\theta_{cr}$ ш Ёх°хэш  тшфр
$$y_1={2p\over3}\cos{\alpha\over3}, \ y_2={2p\over3}\cos{\alpha+2\pi\over3}, \ y_3={2p\over3}\cos{\alpha+4\pi\over3}$$
яЁш $\theta>\theta_{cr}$, уфх
$$\alpha=\arctan\left({2\over r}\sqrt{-{r^2\over4}-{p^3\over27}}\right), \ r(\theta)\neq0.$$
╠юцэю єтшфхЄ№, ўЄю $x_1=y_1+{\theta-1\over3}<0, \
x_2=y_2+{\theta-1\over3}>0, \ x_3=y_3+{\theta-1\over3}>0$ яЁш
$\theta>\theta_{cr}$. ╩Ёюьх Єюую,
$$x_4=y^*+{\theta-1\over3}={1\over2}\left(-1+\sqrt{3}+{1\over3}\sqrt{9+6\sqrt{3}}(3-\sqrt{3})\right)\approx 1.296630263$$
яЁш $\theta=\theta_{cr}$.

─рыхх, шч ЁрсюЄ√ \cite{KRK} (╦хььр 1) ёыхфєхЄ, ўЄю $z_2^{-1}$ ш $z_3^{-1}$  ты ■Єё  Ёх°хэш ьш єЁртэхэш  (\ref{rm}) яЁш $m=2, k=3$. ╚ч тёхую ёърчрээюую ш т ёшыє ЇюЁьєы√ (1.5) шч Єющ цх ЁрсюЄ√ яюыєўшь ёыхфє■∙хх

\textbf{╙ЄтхЁцфхэшх.} ╧єёЄ№ $q=3, k=3, \theta>1$. ╥юуфр ёшёЄхьр єЁртэхэшщ (\ref{pt1})
шьххЄ

1. ┼фшэёЄтхээюх Ёх°хэшх $(1,1)$ яЁш $\theta<\theta_{cr}$;

2. ╥Ёш Ёх°хэш  $(1,1), (z_1,1), (1,z_1)$ яЁш $\theta=\theta_{cr}$;

3. ╤хь№ Ёх°хэшщ $(1,1)$, $(z_2,1)$, $(1,z_2)$, $(z_3,1)$,
$(1,z_3)$, $(z_2,z_2)$, $(z_3,z_3)$  яЁш $\theta>\theta_{cr}$, уфх $z_1=x_4^3, z_2=x_2^3, z_3=x_3^3$.\

\textbf{╟рьхўрэшх 2.} 1. ╙ЄтхЁцфхэшх 1  ты хЄё  ўрёЄэ√ь ёыєўрхь ЄхюЁхь√ 1 шч ЁрсюЄ√ \cite{KRK}. ╟фхё№ ь√ фрыш  тэ√х ЇюЁьєы√ фы  Ёх°хэшщ, ёююЄтхЄёЄтє■∙шї ЄЁрэёы Ўшюээю-шэтрЁшрэЄэ√ь ьхЁрь ├шссёр т ёыєўрх $k=3$.

2. ╧Ёш $\theta=\theta_{cr}$ шьххь $z_1=z_2=z_3$.

3. ╟рьхЄшь, ўЄю Ёх°хэш  $(z_2,z_2)$ ш $(z_3,z_3)$, яюфюсэю ЁрсюЄх \cite{KRK}, яюыєўхэ√ шч Ёх°хэш  тшфр $(1,1,z)$ яЁш $m=1$.

\section{\textbf{╧хЁшюфшўхёъшх ьхЁ√ ├шссёр}}\

╨рёёьюЄЁшь ёыєўрщ $q\geq 3$, Є.х. $\sigma:V\rightarrow\Phi=
\{1,2,3,...,q\}$. ┬ ёшыє ╥хюЁхь√ 2 шьх■Єё  Єюы№ъю
$G^{(2)}_k$-яхЁшюфшўхёъшх ьхЁ√ ├шссёр, ъюЄюЁ√х ёююЄтхЄёЄтє■Є
ёютюъєяэюёЄш тхъЄюЁют $h=\{h_x\in R^{q-1}: \, x\in G_k\}$ тшфр
$$h_x=\left\{%
\begin{array}{ll}
    h, \ \ \ $ хёыш $ |x|-\mbox{ўхЄэю} $,$ \\
    l, \ \ \ $ хёыш $ |x|-\mbox{эхўхЄэю} $.$ \\
\end{array}%
\right. $$
 ╟фхё№ $h=(h_1,h_2,...,h_{q-1}),$ $l=(l_1,l_2,...,l_{q-1}).$
╥юуфр т ёшыє (\ref{f.3}) шьххь:
$$
\left\{%
\begin{array}{ll}
    h_{i}=k\ln{(\theta-1)\exp(l_i) + \sum_{j=1}^{q-1}exp({l_j})+1\over \sum_{j=1}^{q-1}exp({l_j})+\theta},\\[3 mm]
    l_{i}=k\ln{(\theta-1)\exp(h_i) + \sum_{j=1}^{q-1}exp({h_j})+1\over \sum_{j=1}^{q-1}exp({h_j})+\theta},  \\
\end{array}%
i=\overline{1,q-1}. \right.$$\

┬тхфхь ёыхфє■∙шх юсючэрўхэш : $\exp(h_i)=x_i,\ \exp(l_i)=y_i.$
╥юуфр яюёыхфэ■■ ёшёЄхьє єЁртэхэшщ яЁш $i=\overline{1,q-1}$ ьюцэю
яхЁхяшёрЄ№:
\begin{equation}\label{f.4}
\left\{%
\begin{array}{ll}
    x_{i}=\left({(\theta-1)y_i + \sum_{j=1}^{q-1}y_j+1\over \sum_{j=1}^{q-1}y_j+\theta}\right)^k, \\[3 mm]
    y_{i}=\left({(\theta-1)x_i + \sum_{j=1}^{q-1}x_j+1\over \sum_{j=1}^{q-1}x_j+\theta}\right)^k.\\
    \end{array}%
\right.
\end{equation}\

\textbf{╟рьхўрэшх 3.} 1) ╧Ёш $q=2$ ьюфхы№ ╧юЄЄёр ёютярфрхЄ ё
ьюфхы№■ ╚чшэур, ё ъюЄюЁющ ьюцэю ючэръюьшЄ№ё  т ьюэюуЁрЇшш  \cite{R}.

2) ┬ ёыєўрх $k=2, \ q=3$ ш $J<0$ с√ыю фюърчрэю, ўЄю эр
шэтрЁшрэЄэюь ьэюцхёЄтх $I=\{(x_1,x_2,y_1,y_2)\in R^4: x_1=x_2, \
y_1=y_2\}$ тёх $G_k^{(2)}-$яхЁшюфшўхёъшх ьхЁ√ ├шссёр  ты ■Єё 
ЄЁрэёы Ўшюээю-шэтрЁшрэЄэ√ьш (ёь. \cite{RK}).

3) ┬ ёыєўрх $k\geq 1, \ q=3$ ш $J>0$ с√ыю фюърчрэю, ўЄю тёх
$G_k^{(2)}-$яхЁшюфшўхёъшх ьхЁ√ ├шссёр  ты ■Єё 
ЄЁрэёы Ўшюээю-шэтрЁшрэЄэ√ьш (ёь. \cite{RK}).

╚ч ЁрсюЄ√ \cite{KhR} шчтхёЄэю, ўЄю ьэюцхёЄтю
$$I_m=\{z=(u,v)\in R^{q-1}\times R^{q-1}: x_i=x, \ y_i=y,
i=\overline{1,m}, \ x_i=y_i=1, i=\overline{m+1,q-1}\};$$ Є.х.
$u=(\underbrace{x,x,...,x}_m,1,1,...,1), \
v=(\underbrace{y,y,...,y}_m,1,1,...,1)$,
 ты хЄё  шэтрЁшрэЄэ√ь юЄэюёшЄхы№эю юЄюсЁрцхэш  $W:R^{q-1}\times R^{q-1} \rightarrow
R^{q-1}\times R^{q-1},$ юяЁхфхы хьюх ёыхфє■∙шь юсЁрчюь:
$$
\left\{%
\begin{array}{ll}
    x_{i}^{'}=\left({(\theta-1)y_i + \sum_{j=1}^{q-1}y_j+1\over \sum_{j=1}^{q-1}y_j+\theta}\right)^k \\[3 mm]
    y_{i}^{'}=\left({(\theta-1)x_i + \sum_{j=1}^{q-1}x_j+1\over \sum_{j=1}^{q-1}x_j+\theta}\right)^k.\\
    \end{array}%
\right.$$

┬ ёыєўрх $I_m$ ёшёЄхьр єЁртэхэшщ (\ref{f.4}) шьххЄ
ёыхфє■∙шщ тшф:
\begin{equation}\label{f.5}
\left\{%
\begin{array}{ll}
    x=\left({\theta y +(m-1)y+(q-m)\over \theta +m y+(q-m-1)}\right)^k \\[3 mm]
    y=\left({\theta x +(m-1)x+(q-m)\over \theta +m x+(q-m-1)}\right)^k\\
    \end{array}%
\right.
\end{equation}\
 шыш
\begin{equation}\label{f.6}
\left\{%
\begin{array}{ll}
    x=f(y) \\
    y=f(x), \\
    \end{array}%
\right. \texttt{уфх}\ f(x)=\left({\theta x +(m-1)x+(q-m)\over \theta +m
x+(q-m-1)}\right)^k,
\end{equation}.

\textbf{╟рьхўрэшх 4.} 1.(ёь.\cite{KhR}) ╧єёЄ№ $\pi\in S_{q-1}$ яхЁхёЄрэютър.
╬яЁхфхышь фхщёЄтшх $\pi$ эр тхъЄюЁ $x=(x_1,x_2,...,x_{q-1})$ ъръ
$\pi (x)=(x_{\pi(1)},x_{\pi(2)},...,x_{\pi(q-1)})$. ╥юуфр
$\pi(A)=\{(\pi x,\pi y): (x,y)\in A \}$, уфх $A=I_m$, Єръцх  ты хЄё  шэтрЁшрэЄэ√ь ьэюцхёЄтюь юЄэюёшЄхы№эю
$W$, эю ёююЄтхЄёЄтє■∙р  ёшёЄхьр єЁртэхэшщ т ёыєўрх $I_m$ ёютярфрхЄ
ё (\ref{f.5}), яю¤Єюьє эх эрЁє°р  юс∙эюёЄш,
ьюцэю ЁрёёьюЄЁхЄ№ $I_m$.

2. ┬ ЁрсюЄх \cite{KhR} с√ыю фюърчрэю ╙ЄтхЁцфхэшх 1, ўЄю яЁш $k\geq 3$,\ $3\leq
q<k+1$,\, $0<\theta<\overline{\theta}_{cr}=\frac{k-q+1}{k+1}<1$ ёшёЄхьр
єЁртэхэшщ (\ref{f.4}) эр шэтрЁшрэЄэюь ьэюцхёЄтх $I_m$ шьххЄ эх ьхэхх ЄЁхї Ёх°хэшщ $x_0<x_1=1<x_2,$ яЁш
$\theta=\overline{\theta}_{cr}$ шьххЄ эх ьхэхх юфэюую Ёх°хэш  $x_1=1$ ш яЁш $\theta >
\overline{\theta}_{cr}$ шьххЄ Єюы№ъю юфэю Ёх°хэшх $x_1=1$.

┬ ёыхфє■∙хщ ЄхюЁхьх ь√ фюърцхь, ўЄю ёшёЄхьр
єЁртэхэшщ (\ref{f.4}) эр $I_m$, фы  эхъюЄюЁюую $m$, яЁш ¤Єшї єёыютш ї шьххЄ Ёютэю ЄЁш Ёх°хэш  $x_0<x_1=1<x_2$ яЁш $0<\theta<\overline{\theta}_{cr}$.

\textbf{╥хюЁхьр 3.} \textit{╧єёЄ№ $\overline{\theta}_{cr}={k-q+1\over k+1}, \ k\geq 3$,\, $q\geq3$, \, $J<0$.  ╥юуфр фы  ьюфхыш ╧юЄЄёр } \textit{яЁш $0<\theta<\overline{\theta}_{cr}$ ёє∙хёЄтє■Є Ёютэю ЄЁш $G_k^{(2)}-$
яхЁшюфшўхёъшх ьхЁ√ ├шссёр, ёююЄтхЄёЄтє■∙шх ёютюъєяэюёЄш шч ьэюцхёЄтр $I_m$ фы  эхъюЄюЁюую $m$. ╧Ёш ¤Єюь юфэр шч эшї  ты хЄё  ЄЁрэёы Ўшюээю-шэтрЁшрэЄэющ, р фЁєушх фтх $G_k^{(2)}-$
яхЁшюфшўхёъшьш (эх ЄЁрэёы Ўшюээю-шэтрЁшрэЄэ√ьш).}\

\textbf{─юърчрЄхы№ёЄтю.} ╟рьхЄшь, ўЄю ёшёЄхьр єЁртэхэшщ (\ref{f.6}) яЁш $x=y$ ш $\theta<1$ шьххЄ хфшэёЄтхээюх Ёх°хэшх, ъюЄюЁюх ёююЄтхЄёЄтєхЄ ЄЁрэёы Ўшюээю-шэтрЁшрэЄэющ ьхЁх ├шссёр, Є.ъ. яЁш $x=y$ шьххь єЁртэхэшх $x=f(x)$, фы  яЁртющ ўрёЄш ъюЄюЁюую т√яюыэ ■Єё : $f(0)=\left({q-m\over \theta+q-m-1}\right)^k>0$, ЇєэъЎш  $f(x)$ ёЄЁюую єс√трхЄ ш юуЁрэшўхэр яЁш $x>0$. ─юърцхь, ўЄю яЁш єёыютш ї ЄхюЁхь√ ёє∙єёЄтє■Є Єюы№ъю фтх $G_k^{(2)}-$ яхЁшюфшўхёъшх (эх ЄЁрэёы Ўшюээю-шэтрЁшрэЄэ√х) ьхЁ√ ├шссёр ш $x\neq y$. ─ы  ¤Єюую сєфхь шчєўрЄ№ єЁртэхэшх $f(f(x))=x$. ╥ръ ъръ ЇєэъЎш  $f(x)$ юсЁрЄшьр яЁш $x>0$, ЁрёёьюЄЁшь яюёыхфэхх єЁртэхэшх т тшфх $f(x)=f^{-1}(x)=g(x)$, уфх
$$g(x)=f^{-1}(x)={q-m-(\theta+q-m-1)\sqrt[k]{x}\over m\sqrt[k]{x}-\theta-m+1}.$$

╥ръ ъръ $f(x)>0$, Єю $g(x)>0$, юЄъєфр шьххь
$$\theta_1=\left({\theta+m-1\over m}\right)^k<x<\left({q-m\over \theta+q-m-1}\right)^k=\theta_2.$$

╨рёёьюЄЁшь ЇєэъЎш■ $h(x)=\ln{f(x)\over g(x)}=\ln f(x)-\ln g(x)$. ╚чєўшь Ёх°хэш  єЁртэхэш  $h(x)=0$. ▀ёэю, ўЄю $x=1$  ты хЄё  Ёх°хэшхь ¤Єюую єЁртэхэш , Є.х. $h(1)=0$. ╧юы№чє ё№ яЁюшчтюфэ√ьш
$$f'(x)={k(\theta-1)(\theta+q-1)f(x)\over [(\theta+m-1)x+q-m](mx+\theta+q-m-1)}$$
ш
$$g'(x)={(\theta-1)(\theta+q-1)g(x)\over k\sqrt[k]{x^{k-1}}(m\sqrt[k]{x}-\theta-m+1)[q-m-(\theta+q-m-1)\sqrt[k]{x}]},$$
сєфхь шьхЄ№
$$h'(x)={f'(x)\over f(x)}-{g'(x)\over g(x)}={(\theta-1)(\theta+q-1)\over k}\cdot{k^2\over [(\theta+m-1)x+q-m](mx+\theta+q-m-1)}-$$
$$-{(\theta-1)(\theta+q-1)\over k}\cdot{1\over \sqrt[k]{x^{k-1}}(m\sqrt[k]{x}-\theta-m+1)(q-m-(\theta+q-m-1)\sqrt[k]{x})}.$$

╬сючэрўшт $\sqrt[k]{x}=t$, яхЁхяш°хь яЁюшчтюфэє■ $h'(x)$ ёыхфє■∙шь юсЁрчюь
$$v(t)=$$
$${(\theta-1)(\theta+q-1)p(t)\over kt^{k-1}[(\theta+m-1)t^k+q-m](mt^k+\theta+q-m-1)(mt-\theta-m+1)[(\theta+q-m-1)t-q+m]},$$
уфх
$$p(t)=m(\theta+m-1)t^{2k}+k^2m(\theta+q-m-1) t^{k+1}-(k^2-1)(\theta^2+(q-2)\theta+2mq-2m^2-q+1)t^k+$$
$$+k^2(\theta+m-1)(q-m)t^{k-1}+(\theta+q-m-1)(q-m).$$
╟рьхЄшь, ўЄю яю ЄхюЁхьх ─хърЁЄр ю яюыюцшЄхы№э√ї Ёх°хэш ї ьэюуюўыхэр $p(t)$ ш чэрўшЄ $h'(x)$ шьххЄ эх сюыхх фтєї яюыюцшЄхы№э√ї Ёх°хэш .

╦хуъю яЁютхЁшЄ№, ўЄю
$$\lim_{x\rightarrow\theta_1}h(x)=-\infty, \ h(1)=0, \ \lim_{x\rightarrow\theta_2}h(x)=+\infty.$$
╬Єё■фр, єЁртэхэшх $h(x)=0$ шьххЄ яю ъЁрщэхщ ьхЁх юфэю Ёх°хэшх $x_0$ яЁш $x<1$ ш їюЄ  с√ юфэю Ёх°хэшх $x_2$ яЁш $x>1$, хёыш $h'(1)<0$. ╚ч ¤Єюую єёыютш 
$$h'(1)={(\theta-1)(\theta+q-1)\over k}\cdot\left[{k^2\over (\theta+q-1)^2}-{1\over (\theta-1)^2}\right]<0$$
т√ЄхърхЄ $\theta<{k-q+1\over k+1}=\overline{\theta}_{cr}$, Є.ъ. ${(\theta-1)(\theta+q-1)\over k}<0$.

╩Ёюьх Єюую,
$$\lim_{x\rightarrow\theta_1}h'(x)=+\infty, \ \lim_{x\rightarrow\theta_2}h'(x)=+\infty$$
яЁш $\theta<\overline{\theta}_{cr}$. ╥юуфр шч $h'(1)<0$ ёыхфєхЄ, ЇєэъЎш  $h(x)$ шьххЄ Ёютэю фтх ъЁшЄшўхёъшх Єюўъш $\xi_1$, $\xi_2$: $\theta_1<\xi_1<1$ ш $1<\xi_2<\theta_2$.
╟эрўшЄ, $h(x)$ тючЁрёЄрхЄ яЁш $\theta_1<x<\xi_1, \ x>\xi_2$ ш єс√трхЄ яЁш $\xi_1<x<\xi_2$ (╨шё.1). ╤ыхфютрЄхы№эю, єЁртэхэшх $h(x)=0$ шьххЄ Єюы№ъю ЄЁш Ёх°хэш  $x_0<x_1=1<x_2$. ╬ёЄрыюё№ яЁютхЁшЄ№ єёыютшх $x\neq y$. ╠√ шьххь Ёх°хэш  $x_0<x_1=1<x_2$. ╥юуфр, Є.ъ. ЇєэъЎш  $f(x)$ ёЄЁюую єс√трхЄ, шч тЄюЁюую єЁртэхэш  ёшёЄхь√ (\ref{f.6}) яюыєўшь, ўЄю $f(x_0)=y_0>f(x_1)=y_1=1>f(x_2)=y_2$. ╚Єръ, фы  ьюфхыш ╧юЄЄёр яЁш єёыютш ї ЄхюЁхь√ ёє∙хёЄтє■Є Єюы№ъю ЄЁш $G_k^{(2)}-$
яхЁшюфшўхёъшх ьхЁ√ ├шссёр $\mu_0, \mu_1, \mu_2$, ёююЄтхЄёЄтє■∙шх Ёх°хэш ь $x_0<x_1=1<x_2$, ёююЄтхЄёЄтхээю, уфх ьхЁр $\mu_1$  ты хЄё  ЄЁрэёы Ўшюээю-шэтрЁшрэЄэющ, р ьхЁ√ $\mu_0, \mu_2$  ты ■Єё  $G_k^{(2)}-$
яхЁшюфшўхёъшьш (эх ЄЁрэёы Ўшюээю-шэтрЁшрэЄэ√ьш). ╥хюЁхьр фюърчрэр.\

\begin{center}
\includegraphics[width=6cm]{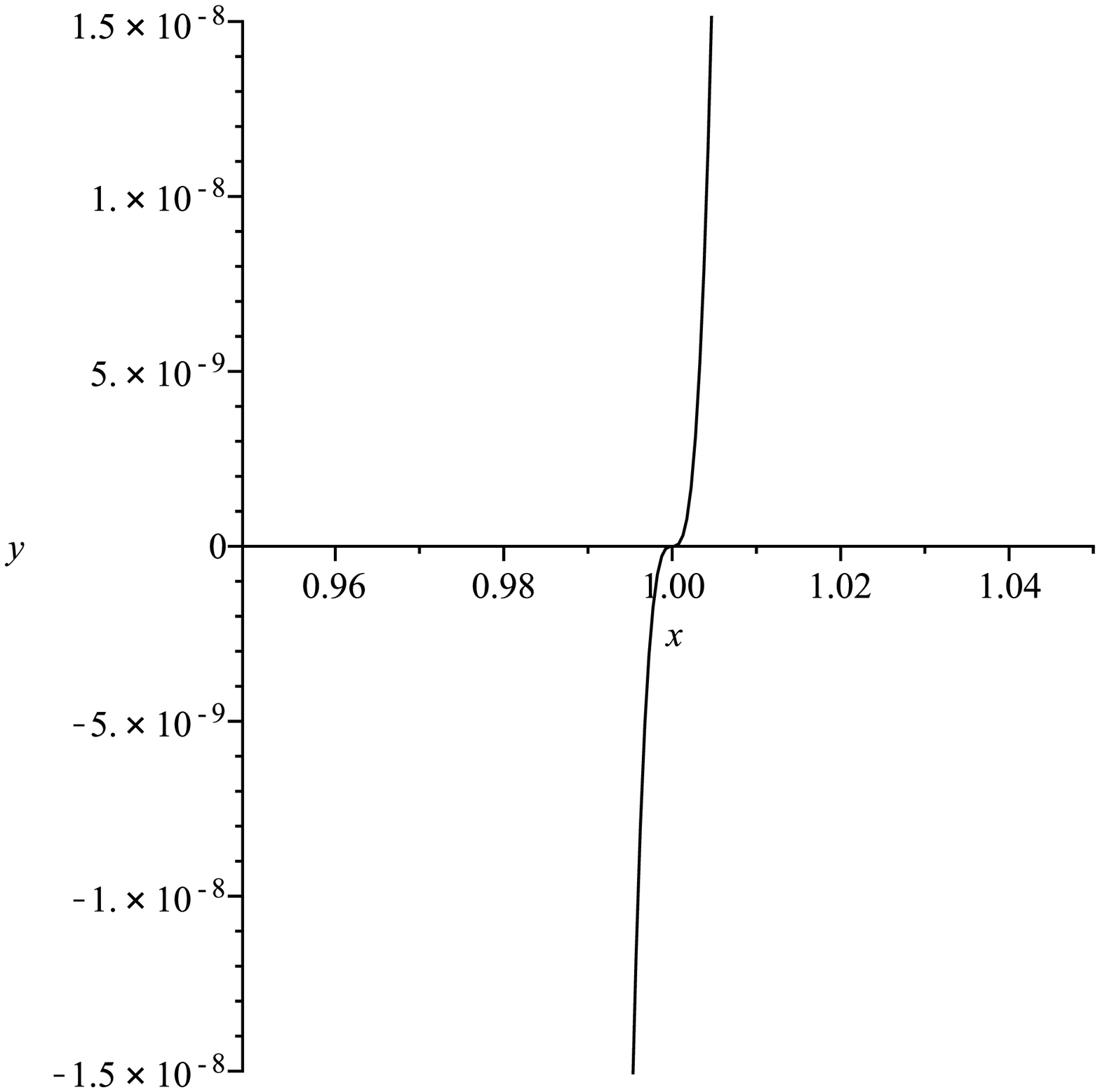} \includegraphics[width=6cm]{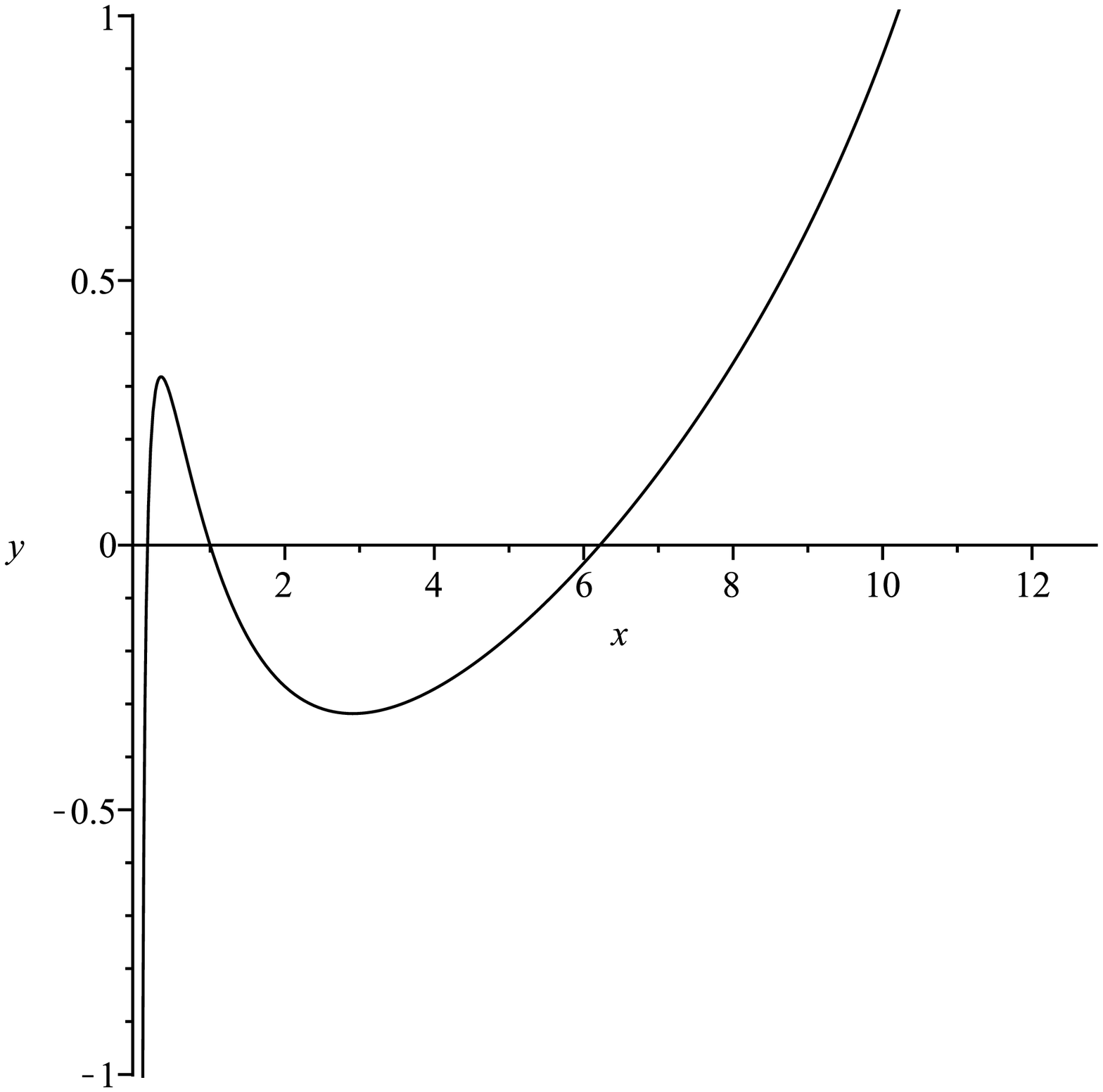}
\end{center}
\begin{center}{\footnotesize \noindent
 ╨шё.~1.
├ЁрЇшъ ЇєэъЎшш $h(x)$ яЁш $k=3, \ q=3, \ \overline{\theta}_{cr}=0.25, \ m=1$ (ёыхтр) ш яЁш $k=5, \ q=4, \ \theta=0.2, \ m=2$ (ёяЁртр).}\\
\end{center}

┬ ёшыє ╥хюЁхь√ 3 шч \cite{KhR}, ёяЁртхфыштр ёыхфє■∙р 

\textbf{╥хюЁхьр 4.} {─ы  ьюфхыш ╧юЄЄёр яЁш $k\geq 3$,\ $3\leq
q<k+1$ ш $0<\theta<\overline{\theta}_{cr}$ эр $\bigcup_{m=1}^{q} I_m$ ёє∙хёЄтє■Є Ёютэю
$$2\cdot (2^{q}-1)$$
$G_k^{(2)}-$ яхЁшюфшўхёъшї (эх ЄЁрэёы Ўшюээю-шэтрЁшрэЄэ√ї) ьхЁ
├шссёр}.\

\textbf{┴ыруюфрЁэюёЄш.} └тЄюЁ√ т√Ёрцр■Є уыєсюъє■ яЁшчэрЄхы№эюёЄ№ яЁюЇхёёюЁє ╙.└.╨ючшъютє чр яюыхчэ√х ёютхЄ√.


\begin{thebibliography}{99}

\bibitem{6} ╒.-╬. ├хюЁуш. \textit{├шссёютёъшх ьхЁ√ ш Їрчют√х яхЁхїюф√}. - ╠.: ╠шЁ, 1992.

 \bibitem{Pr} C. J. Preston. \textit{Gibbs States on Countable Sets}. - Cambridge Tracts Math., 68, Cambridge
Univ. Press, Cambridge, 1974.

\bibitem{Si} ▀. ├. ╤шэрщ. \textit{╥хюЁш  Їрчют√ї яхЁхїюфют. ╤ЄЁюушх Ёхчєы№ЄрЄ√}. - M.: ═рєър, 1980.

\bibitem{Ga8} ═.═. ├рэшїюфцрхт. \textit{╥╠╘}, \textbf{85}: 2 (1990), 163--175.

\bibitem{GN} ═.═. ├рэшїюфцрхт. \textit{─└═ ╨╙ч},  6-7  (1992), 4--7.

\bibitem{R} U.A. Rozikov. \textit{Gibbs measures on Cayley
trees}. World Scientific.-2013.

\bibitem{Ga13} N.N. Ganikhodjaev, U.A. Rozikov. \textit{Lett. Math. Phys}.  \textbf{75}: 2  (2006), 99-Ц109.

\bibitem{RK} ╙.└. ╨ючшъют, ╨.╠. ╒ръшьют. \textit {╥╠╘}, \textbf{175}:
2 (2013), 300--312.

\bibitem{KhR1} ╒ръшьют ╨.╠. ╬ ёє∙хёЄтютрэшш яхЁшюфшўхёъшї ьхЁ ├шссёр фы  ьюфхыш
╧юЄЄёр эр фхЁхтх ╩¤ыш, \textit{╙чсхъёъшщ ьрЄхьрЄшўхёъшщ цєЁэры}, No 3, 2014, 134--142.

\bibitem{KhR} R. M. Khakimov. New periodic Gibbs measures for $q$-state Potts
model on a Cayley tree. \textit{Journal of Siberian Federal
University. Mathematics and Phisics.} 2014, 7(3), p.297-304.

\bibitem{KRK} C. K\"ulske, U. A. Rozikov, R. M. Khakimov. Description of all translation-invariant (splitting) Gibbs measures
for the Potts model on a Cayley tree. \textit{Jour. Stat. Phys.} \textbf{156}(1) (2014), 189-200.

\bibitem{KR} C. K\"ulske, U.A. Rozikov, \textit{ Fuzzy transformations and extremality of Gibbs measures
for the Potts model on a Cayley tree}, arXiv:1403.5775v1
[math-ph].




\end{thebibliography}
\end{document}